\documentclass[conference]{IEEEtran}
\usepackage{cite}
\usepackage{amsmath,amssymb,amsfonts}
\usepackage{graphicx}
\usepackage{textcomp}
\usepackage{url}
\usepackage{hyperref}
\usepackage{xcolor}
\usepackage{multirow}
\usepackage{booktabs}
\usepackage{tikz}
\usepackage{colortbl}
\usetikzlibrary{positioning, arrows.meta, shapes.geometric, calc, fit, backgrounds}
\usepackage{tabularx}
\usepackage[caption=false,font=footnotesize]{subfig}
\def\BibTeX{{\rm B\kern-.05em{\sc i\kern-.025em b}\kern-.08em
    T\kern-.1667em\lower.7ex\hbox{E}\kern-.125emX}}
\begin{document}

\title{Automated Parameter Sensitivity Ranking in Audio Synthesis\\
{}
}

\author{\IEEEauthorblockN{Oğulcan KİBAR}
\IEEEauthorblockA{\textit{Graduate School \& Burhaniye Vocational School} \\
\textit{Bahçeşehir University \& Balikesir University}\\
İstanbul, Turkey \& Balıkesir, Türkiye\\
0000-0001-6880-6404}
\and
\IEEEauthorblockN{Barış ÖZCAN}
\IEEEauthorblockA{\textit{Computer Engineering Department} \\
\textit{Bahçeşehir University}\\
İstanbul, Turkey \\
0000-0003-4667-8047}
}

\maketitle

\begin{abstract}
Software synthesizers expose dozens of coupled controls, many of them redundant or perceptually inert at typical operating points. This paper presents a two-stage, measure-and-validate framework that ranks parameters by their influence on the audio output and then checks that ranking against an independent, variance-based experiment. Every rendered clip is normalized to $-23$\,LUFS and described by five spectral fingerprints, so that gain shifts cannot masquerade as timbral change. The Sweep Test scores each parameter by spectral deviation over a 101-step sweep; the Sobol Test scores it by summed per-cell variance over a 128-point Sobol excitation. Across five architecturally distinct synthesizers (TorchSynth, Pyo, Pedalboard, Csound, and FluidSynth) spanning 48 to 76 parameters, the two tests agree at Spearman $\rho = 0.82\text{--}0.96$ ($p$-values $\le 7.2\times10^{-15}$) on the four engines that render deterministically, and at $\rho = 0.32$ on the one with process-level render noise—a diagnostic of that engine rather than of the method. The strong correlations survive reducing the Sobol budget from 128 to 32 points, and the complete analysis costs a one-time offline budget of about 66,000 renders.
\end{abstract}

\begin{IEEEkeywords}
audio synthesis, variance-based sensitivity analysis, Sobol sequences, perturbation analysis, parameter ranking
\end{IEEEkeywords}

% ============================================================
\section{Introduction}
% ============================================================

A musician who opens a modern software synthesizer faces a wall of several dozen controls, and behind it a practical question this paper answers quantitatively: \emph{which of these knobs actually matter for the sound?} Systems such as TorchSynth \cite{turian2021billionaudiosoundsgpuenabled} expose 76 or more synthesis parameters per note; many are redundant or perceptually inert at typical operating points, and equifinality obscures the relationship between individual controls and the audible output.

Answering the question is the motivation for performing a sensitivity analysis of a synthesizer in the first place. Practically, a reliable ranking tells an instrument designer which controls deserve front-panel space, a sound designer where to spend exploration effort, and an automated search which dimensions can be frozen without losing sonic territory. Scientifically, knowing which parameters influence the output is the prerequisite for any principled dimension reduction of a control space—and a ranking is only trustworthy if it can be checked. The rationale of this study is therefore twofold: to measure the sensitivity of every parameter of a synthesizer treated strictly as a black box, and to validate the resulting ranking with a second, methodologically independent experiment.

Two audio-specific obstacles stand in the way. Some parameters change only loudness, so any distance computed on unnormalised audio is dominated by gain effects and mistakes them for timbral influence; and audio is a time series, where a microscopic phase shift produces an enormous sample-wise difference, so comparison must happen in time--frequency representations rather than in the waveform domain \cite{steinmetz2020auraloss}.

This paper contributes a two-stage architecture that addresses both obstacles and validates itself. Conditioning normalises every clip to $-23$\,LUFS \cite{EBUR128_2020, itur_bs1770_5_2023}, removing the gain confound at the source, and five complementary fingerprints—log-STFT, log-mel, MFCC \cite{davis1980comparison}, RMS envelope, and spectral centroid \cite{peeters2004large}—make the comparison meaningful. The Sweep Test assigns each parameter a local score $S_i$ from a 101-step sweep against the mid-range reference clip; the Sobol Test independently assigns an output-variance score $V_i$ from 128 quasi-uniform excitations \cite{SOBOL196786, joe2008constructing}; the Spearman rank correlation \cite{spearman2010proof} between the two rankings validates the framework. Across five architecturally unrelated engines the tests agree at $\rho = 0.96$, $0.92$, $0.87$, and $0.82$ where rendering is deterministic and at $0.32$ where it is not, both select the same most-sensitive parameter on Pedalboard and Csound, and the three strongest correlations change by at most $0.006$ when the Sobol budget drops to 32 points. The scope of the contribution is the ranking and its validation: applying it to reduce a control space, and confirming that quality is preserved, is identified as the next step rather than reported here.

% ============================================================
\section{Related Work}
\label{sec:related}
% ============================================================

\emph{Sensitivity analysis.} Two traditions dominate. Local, one-at-a-time designs perturb a single input while the others are held fixed; they are cheap and interpretable \cite{morris1991factorial} but probe a single operating point and ignore interactions \cite{SALTELLI20101508}. Global, variance-based methods instead attribute output variance to the inputs \cite{SOBOL2001271, saltelli:2008}, decomposing it into first-order and total-order contributions estimated from sampling plans whose cost grows with the number of inputs, and are driven by constructed designs \cite{mckay1979comparison} or low-discrepancy sequences \cite{SOBOL196786, Caflisch_1998}. Such sequences are often used deterministically, and that determinism is known to interact with index estimation in high-dimensional designs \cite{Sun2021}. This work differs in taking the Sobol sequence as a per-parameter \emph{excitation design} rather than as the sampling plan of an index estimator: it computes no first- or total-order index, trading interaction information for a cost linear in the number of parameters and for two tests whose rankings can be compared against each other.

\emph{Comparing synthesized audio.} Any sensitivity measure on audio needs a representation, and the choice is not neutral. Waveform-domain differences are unusable because a microscopic phase shift produces an enormous sample-wise difference \cite{steinmetz2020auraloss}, so comparison is normally made on time--frequency maps or on scalar descriptors such as MFCCs \cite{davis1980comparison} and the spectral centroid \cite{peeters2004large}. Work comparing synthesized sounds usually reduces the comparison to a single spectral distance, yet benchmarks show that such distances can order even elementary pitch differences incorrectly \cite{turian2020imsorrylossspectrallybased}, and spectrogram-domain comparison misses structure on longer time scales than one frame captures \cite{vahidi2023mesostructuresspectrogramlossdifferentiable}. Studies of how evaluation metrics themselves respond to parameter changes reach a compatible conclusion \cite{gupta2022parametersensitivitydeepfeaturebased}, and recent parameter-recovery work accordingly optimises several complementary descriptors at once \cite{bogdan2026instrumentalautomaticsynthesizerparameter}. This work differs in scoring every excitation on five complementary fingerprints and aggregating them symmetrically on both sides of the framework, so that no single representation decides the ranking.

\emph{Sensitivity analysis in audio.} Sensitivity has been studied inside audio, usually with a local criterion and a narrow output description: the geometric sensitivity of a saxophone resonator, for instance, is derived analytically and read from modal frequencies and harmonicity \cite{szwarcberg2025geometricsensitivitymodalparameters}. Parameter-space dimensionality is repeatedly identified as the bottleneck in tasks that search these spaces, such as sound matching \cite{salimi2025evaluatingsoundsimilaritymetrics} and the resynthesis of natural sounds \cite{hagiwara2022modelinganimalvocalizationssynthesizers}, and studies of high-dimensional configuration spaces show that most inputs contribute negligible output variance \cite{novello2022goalorientedsensitivityanalysishyperparameters}. This work differs in treating the instrument strictly as a black box mapping $[0,1]^N$ to audio, requiring no access to its internal structure, and in validating the resulting ranking with a second, methodologically independent test rather than reporting it unchecked.

% ============================================================
\section{Proposed Methodology}
% ============================================================

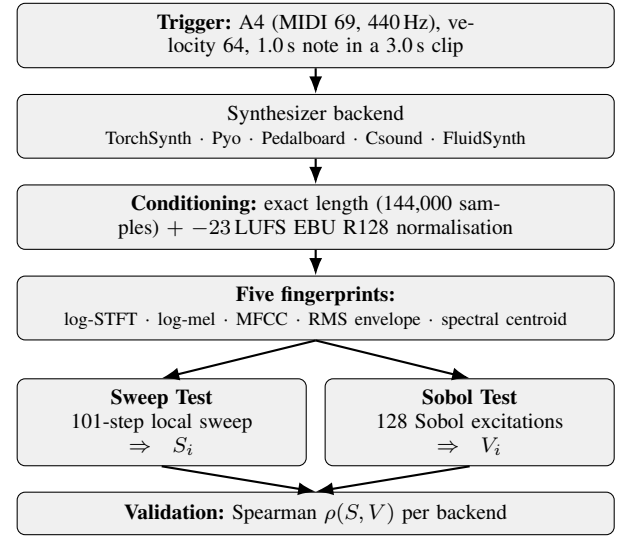
\begin{figure}[t]
\centering
\begin{tikzpicture}[
  node distance=0.34cm,
  pbox/.style={draw, rectangle, rounded corners=3pt,
               fill=gray!12, align=center, font=\footnotesize,
               text width=0.86\columnwidth, inner sep=4pt},
  hbox/.style={draw, rectangle, rounded corners=3pt,
               fill=gray!12, align=center, font=\footnotesize,
               text width=0.40\columnwidth, inner sep=4pt},
  parr/.style={-Latex, thick}
]
  \node[pbox] (trig)  {\textbf{Trigger:} A4 (MIDI 69, 440\,Hz), velocity 64, 1.0\,s note in a 3.0\,s clip};
  \node[pbox, below=of trig] (rend)  {Synthesizer backend\\
    {\scriptsize TorchSynth $\cdot$ Pyo $\cdot$ Pedalboard $\cdot$ Csound $\cdot$ FluidSynth}};
  \node[pbox, below=of rend] (cond)  {\textbf{Conditioning:} exact length (144{,}000 samples) $+$ $-23$\,LUFS EBU~R128 normalisation};
  \node[pbox, below=of cond] (feat)  {\textbf{Five fingerprints:}\\
    {\scriptsize log-STFT $\cdot$ log-mel $\cdot$ MFCC $\cdot$ RMS envelope $\cdot$ spectral centroid}};
  \node[hbox, below=0.5cm of feat, xshift=-0.23\columnwidth] (t1)
    {\textbf{Sweep Test}\\ 101-step local sweep\\ $\Rightarrow S_i$};
  \node[hbox, below=0.5cm of feat, xshift=0.23\columnwidth] (t2)
    {\textbf{Sobol Test}\\ 128 Sobol excitations\\ $\Rightarrow V_i$};
  \node[pbox, below=2.0cm of feat] (val)
    {\textbf{Validation:} Spearman $\rho(S, V)$ per backend};
  \draw[parr] (trig) -- (rend);
  \draw[parr] (rend) -- (cond);
  \draw[parr] (cond) -- (feat);
  \draw[parr] (feat.south) -- (t1.north);
  \draw[parr] (feat.south) -- (t2.north);
  \draw[parr] (t1.south) -- (val.north);
  \draw[parr] (t2.south) -- (val.north);
\end{tikzpicture}
\caption{Overview of the framework.}
\label{fig:pipeline}
\end{figure}

Figure~\ref{fig:pipeline} summarizes the two-stage measure-and-validate architecture. In this framework, both sensitivity tests operate on identically conditioned audio clips and identical spectral fingerprints, differing only in their parameter excitation strategy and scoring method. The complete pipeline comprises standardized audio triggering and loudness conditioning, spectral fingerprint extraction, the two sensitivity scoring tests, and a final Spearman rank-correlation validation.

Notation is fixed throughout the paper: $p$ denotes a normalised parameter value in $[0,1]$ and $p_i$ the value of the $i$-th parameter, $S_i$ and $V_i$ are the two sensitivity scores, and $\rho$ is the Spearman rank-correlation coefficient. Statistical significance is always written explicitly as a ``$p$-value'' and never as a bare~$p$, so that parameter values, correlation coefficients, and significance levels remain distinguishable.

\subsection{Experimental Setup and Signal Conditioning}
\label{sec:conditioning}
All experiments use a standardised single-note trigger: pitch A4 (MIDI note 69, $f_0 = 440$\,Hz), velocity 64, and a 1.0\,s note-on inside a 3.0\,s clip rendered at 48\,kHz (144{,}000 samples), leaving a 2.0\,s tail in which release and decay behaviour remains observable. All synthesis parameters are normalised to $[0,1]$, and the mid-range baseline fixes every background parameter at $0.5$.

Before any analysis, every rendered clip passes a mandatory conditioning stage. The clip is first padded to exactly 144{,}000 samples, a requirement for the per-cell variance computation of the Sobol Test. It is then normalised to $-23$\,LUFS integrated loudness following EBU~R128 \cite{EBUR128_2020}, with the integrated-loudness measurement defined by ITU-R BS.1770-5 \cite{itur_bs1770_5_2023}; clips whose loudness falls below the $-70$\,LUFS gating threshold are passed through unmodified, so silence remains silence. This normalisation is the framework's answer to the gain confound: louder output reliably registers as different even when its character is unchanged, so parameters that only make the output louder or quieter could otherwise masquerade as timbre-altering controls. After matching, the subsequent statistics respond to structural, spectral, and temporal change only. One consequence is documented rather than hidden: a parameter that mutes the output entirely at one end of its range, like a master volume set to $0$, still registers as influential, because silence versus sound is a genuine structural difference that survives conditioning through the gate.

\subsection{From Waveforms to Comparable Representations}
\label{sec:fingerprints}
Every conditioned clip is described by five fingerprints, computed independently of one another:
\begin{itemize}
  \item \textbf{Log-STFT}: $\log_{10}(|\mathrm{STFT}|^2 + 10^{-8})$, with a $1024$-point FFT, a $256$-sample hop and a Hann window.
  \item \textbf{Log-mel}: the power spectrogram pooled through $128$ triangular mel filters spanning $20\text{--}8000$\,Hz.
  \item \textbf{MFCC}: the first $40$ coefficients of the orthonormal discrete cosine transform of the log-mel map \cite{davis1980comparison}.
  \item \textbf{RMS envelope}: frame-wise RMS with the same $1024/256$ framing.
  \item \textbf{Spectral centroid}: the amplitude-weighted mean frequency per frame \cite{peeters2004large}.
\end{itemize}

The first three are two-dimensional maps, the last two one-dimensional tracks. They carry incompatible units, so both tests combine them by the same two-step recipe: each fingerprint's raw responses are min--max rescaled across the parameters of the backend, and the five rescaled values are averaged with equal weight. The recipe is written out for the Sweep Test in \eqref{eq:si} and applied identically to the Sobol Test variances in Section~\ref{sec:test2}. The supplementary material states what each fingerprint contributes and why the weights are held fixed.

\subsection[Sweep Test]{Sweep Test ($S_i$)}
\label{sec:test1}
The Sweep Test measures how strongly the sound departs from its mid-range state when a single parameter is moved step by step across its full range. For each analysed parameter $p_i$, all other parameters are frozen at $0.5$ and $p_i$ takes $G = 101$ uniform values $0.00, 0.01, \dots, 1.00$. The sweep's own centre clip ($p_i = 0.5$, identical to the global baseline) serves as the reference. For each of the remaining $100$ clips, five scalar deviations are computed against the reference, and averaging over the sweep yields a raw per-metric response
\begin{equation}
\bar{\delta}_m^{(i)} = \frac{1}{G-1} \sum_{g \ne \mathrm{ref}} \delta_m\!\left( \mathcal{G}(p^{(g)}),\, \mathcal{G}(p^{(\mathrm{ref})}) \right).
\label{eq:rawdev}
\end{equation}
Every symbol in \eqref{eq:rawdev} is defined as follows: $i$ indexes the analysed parameter and $p_i$ is that parameter; $G = 101$ is the number of grid points of the sweep; $g$ indexes those grid points and $\mathrm{ref}$ denotes the centre point at which $p_i = 0.5$; $p^{(g)} \in [0,1]^N$ is the complete parameter vector whose $i$-th entry carries the $g$-th grid value while all remaining entries stay at $0.5$; $\mathcal{G}(\cdot)$ is the synthesis-plus-conditioning pipeline of Section~\ref{sec:conditioning}, mapping a parameter vector to a conditioned audio clip; $m$ indexes the five fingerprint metrics of Section~\ref{sec:fingerprints}, collected in the set $\mathcal{M}$ with $|\mathcal{M}| = 5$; $\delta_m(\cdot,\cdot)$ is the scalar deviation between two clips under metric $m$—the absolute difference for the two scalar summaries (global RMS and mean spectral centroid) and the $L_2$ (Frobenius) norm of the map difference for the log-STFT, log-mel, and MFCC maps; and $\bar{\delta}_m^{(i)} \ge 0$ is the resulting mean response of metric $m$ to parameter $i$, averaged over the $G - 1 = 100$ non-reference clips.

Two safeguards make \eqref{eq:rawdev} robust. First, a \emph{render-noise gate}: the baseline configuration is rendered twice, the deviation between the two renders defines a per-metric noise floor, and any metric whose sweep deviations never exceed three times that floor is treated as unresponsive ($\bar{\delta}_m^{(i)} := 0$). On deterministic engines the floor is exactly zero and the gate is inert; on engines with render noise it prevents pure noise from being mistaken for sensitivity. Second, the aggregation across metrics is \emph{magnitude-preserving}: each metric's raw responses are min--max normalised \emph{across the parameters} of the backend, and the final score is their mean,
\begin{equation}
S_i = \frac{1}{|\mathcal{M}|} \sum_{m \in \mathcal{M}}
\frac{\bar{\delta}_m^{(i)} - \min_j \bar{\delta}_m^{(j)}}
     {\max_j \bar{\delta}_m^{(j)} - \min_j \bar{\delta}_m^{(j)}} \;\; \in [0, 1].
\label{eq:si}
\end{equation}
In \eqref{eq:si}, $j$ ranges over all $N$ analysed parameters of the backend, so the fraction places parameter $i$'s response on a $[0,1]$ scale relative to its peers, separately for each metric, and the outer sum averages the five rescaled responses with equal weights of $1/|\mathcal{M}| = 1/5$. The Sweep Test is a one-at-a-time design in the spirit of \cite{morris1991factorial}.

\subsection[Sobol Test]{Sobol Test ($V_i$)}
\label{sec:test2}
The Sobol Test measures the variance that appears in the output when a parameter receives quasi-random shocks across its range. With the others again frozen at $0.5$, $p_i$ takes $K = 128$ values from the deterministic one-dimensional Sobol sequence \cite{SOBOL196786, joe2008constructing}. The sequence is used unscrambled, so every parameter of every backend receives the identical 128 input values and the experiment regenerates exactly.

The 128 conditioned clips are converted to the five fingerprint maps of Section~\ref{sec:fingerprints}, the 128 instances of each map are stacked, and the variance of every cell is computed across the stack and summed:
\begin{equation}
V_m^{(i)} = \sum_{\mathrm{cells}\ c} \operatorname{Var}_{k = 1..K}\!\left[ \mathcal{F}_m^{(i,k)}(c) \right].
\label{eq:vraw}
\end{equation}
The symbols of \eqref{eq:vraw} are defined as follows: $c$ indexes the cells of fingerprint $m$—a single time--frequency bin for the three two-dimensional maps, and a single frame for the two one-dimensional tracks; $k = 1, \dots, K$ indexes the Sobol clips; $\mathcal{F}_m^{(i,k)}(c)$ is the value of cell $c$ of fingerprint $m$ computed from the $k$-th clip of parameter $i$; and $\operatorname{Var}$ is the unbiased sample variance across the $K$ stacked maps (denominator $K - 1$).

\begin{figure}[t]
  \centering
  \includegraphics[width=0.68\columnwidth]{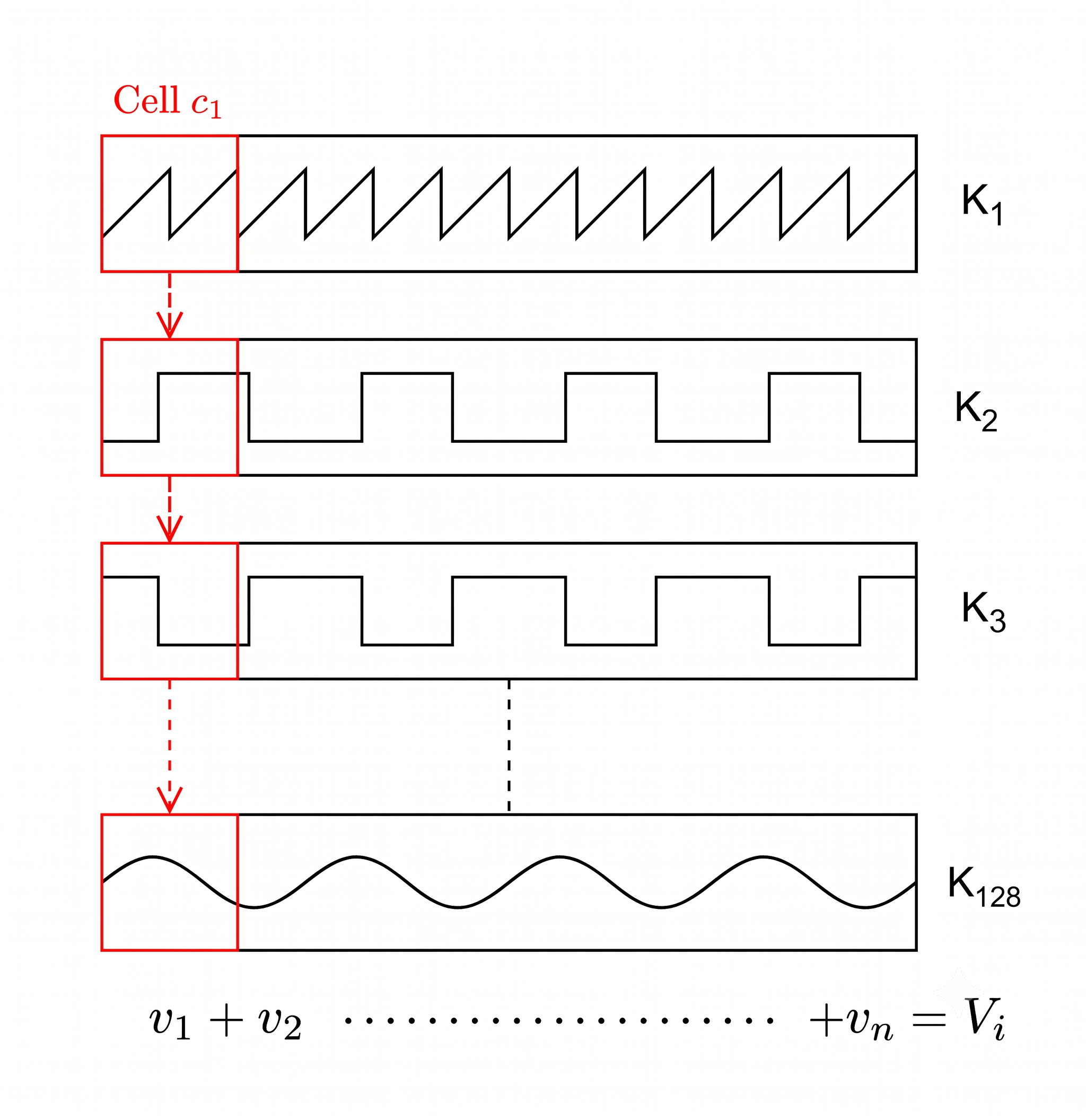}
  \caption{Per-cell variance of the Sobol Test: each row is one of the $K$ conditioned clips, variance is taken down each cell $c$, and the sum over cells gives $V_i$.}
  \label{fig:sobol_test_output-variance}
\end{figure}

The per-cell evaluation procedure is illustrated in Fig.~\ref{fig:sobol_test_output-variance}. For a given parameter $p_i$, the $K = 128$ conditioned clips are mapped to their corresponding time--frequency fingerprints and stacked. As the diagram shows, the sample variance is evaluated vertically across the $K$ stacked instances for each individual cell $c$, and the scalar sum over all cells yields the raw score, which becomes $V_i$ once the variance floor described below is subtracted. A strong parameter produces 128 mutually different maps and a large summed variance; a weak one produces near-copies and a variance near zero.

 A two-render baseline procedure, identical to the Sweep Test gate, provides a variance floor that is subtracted from each raw score. The log-STFT variance is the headline score $V_i$, the other four are reported as transparency columns, and an aggregated score $V_i^{\mathrm{agg}}$ mirrors \eqref{eq:si} exactly: each of the five per-fingerprint variances is min--max normalised across the parameters of the backend and the five values averaged with the same weights. Section~\ref{sec:val_results} reports both.

\subsection{System Validation}
\label{sec:validation}
The two tests share their conditioning and fingerprints but differ in everything else: a uniform grid against quasi-random excitation, a distance-from-reference score against a dispersion score. If both produce the same ordering, that ordering is a property of the synthesizer rather than of either method. Agreement is quantified by the Spearman rank correlation \cite{spearman2010proof} between the $S$ and $V$ lists of each backend, with tied scores handled by the average-rank convention; parameters with large rank gaps are flagged for inspection rather than discarded. The supplementary material states why a rank statistic is the appropriate choice here.

% ============================================================
\section{Experimental Setup}
\label{sec:setup}
% ============================================================

All five synthesizers accept a normalized parameter vector $p \in [0,1]^{N}$, ensuring an identical analysis loop across every backend without backend-specific branching. The evaluation engines include:
\begin{itemize}
    \item \textbf{TorchSynth} ($N = 76$): A PyTorch reimplementation of the TorchSynth Voice \cite{turian2021billionaudiosoundsgpuenabled} with GPU-accelerated batch rendering.
    \item \textbf{Pyo} ($N = 56$): A Python DSP server \cite{Belanger2016} featuring two oscillators, a filter, envelopes, LFOs, reverb, chorus, delay, and distortion.
    \item \textbf{Pedalboard} ($N = 56$): A DSP processing chain built on Spotify's Pedalboard library \cite{sobot_peter_2023_7817838} utilizing the same module set as Pyo.
    \item \textbf{Csound} ($N = 52$): An orchestra instrument compiled at runtime from a CSD template \cite{Lazzarini2016CsoundAS}.
    \item \textbf{FluidSynth} ($N = 48$ analyzed of $50$): An SF2 SoundFont renderer\footnote{\url{https://www.fluidsynth.org}} driven by MIDI controllers and SF2 generator overrides using the General MIDI Grand Piano preset.
\end{itemize}

Two FluidSynth controls are level properties rather than synthesis parameters and are pinned to $64/127$ and excluded from the analysis: \texttt{note.velocity} and \texttt{cc.volume}.

The Sweep Test renders $N \times 101$ clips per backend and the Sobol Test renders $N \times 128$, for a total of $29{,}088$ and $36{,}864$ renders respectively across all five engines. All random seeds are fixed at 42; the Sobol sequence is used unscrambled and is therefore fully deterministic. Every conditioned Sobol Test clip is additionally exported as a 16-bit WAV file. That corpus, the result tables and run logs, and a per-parameter reference listing every analysed control with both scores, both ranks and the rank gap are provided as supplementary material through the repository linked in Section~\ref{sec:conclusion}; no table of parameter ranges is needed, because every parameter is normalised to $[0,1]$ by construction. TorchSynth, Pedalboard, and Csound render bit-identically, as does Pyo once a discarded warm-up render absorbs a one-off initialisation transient. FluidSynth is the exception, and in an instructive way: repeated renders of an unchanged parameter vector are reproducible within some measurement processes and not within others. Its Sweep Test process recorded a non-zero render-to-render floor—$218$ on the log-STFT map, identical to three significant figures in two runs five days apart, so the discrepancy is repeatable rather than stochastic—whereas the Sobol Test floor was exactly zero in every process and at all three budgets. The two safeguards are therefore not symmetric in practice: the Sweep Test gate had a real floor to act on, while the Sobol Test floor subtracted nothing because that test never encountered the variation, an asymmetry Section~\ref{sec:val_results} shows to be the direct cause of the two tests' disagreement here.

% ============================================================
\section{Results}
\label{sec:results}
% ============================================================

\subsection{Sweep Test Rankings}
Table~\ref{tab:top5} lists the five highest-scoring parameters per backend, and the left half of Table~\ref{tab:top15} shows the full top-15 for TorchSynth; ranks throughout are average ranks, so parameters that tie share a fractional rank. The rankings are physically plausible throughout. TorchSynth is dominated by the release stages of its amplitude envelopes, followed by the tuning and modulation depth of its second oscillator. Pyo is likewise led by the release of its amplifier envelope ($S_i = 0.95$), with its filter-type selector and filter resonance next. Pedalboard and Csound are each led by a discrete filter-mode switch at the normalisation ceiling: flipping a filter between low-pass and high-pass is the single most drastic spectral change either engine can produce with one control. FluidSynth's nominal leader, the sustain level of its SF2 modulation envelope ($S_i{=}1.0$), is an artefact of that engine's process-level render noise and is dissected in Section~\ref{sec:val_results}; its highest genuinely responsive control is the expression pedal controller, the remaining gain-family parameter after the volume controller was pinned as a level property at $p{=}0$ it mutes the note entirely, and silence versus sound is a structural difference that loudness normalisation deliberately preserves (Section~\ref{sec:conditioning}).

\begin{figure}[t]
  \centering
  \includegraphics[width=\columnwidth]{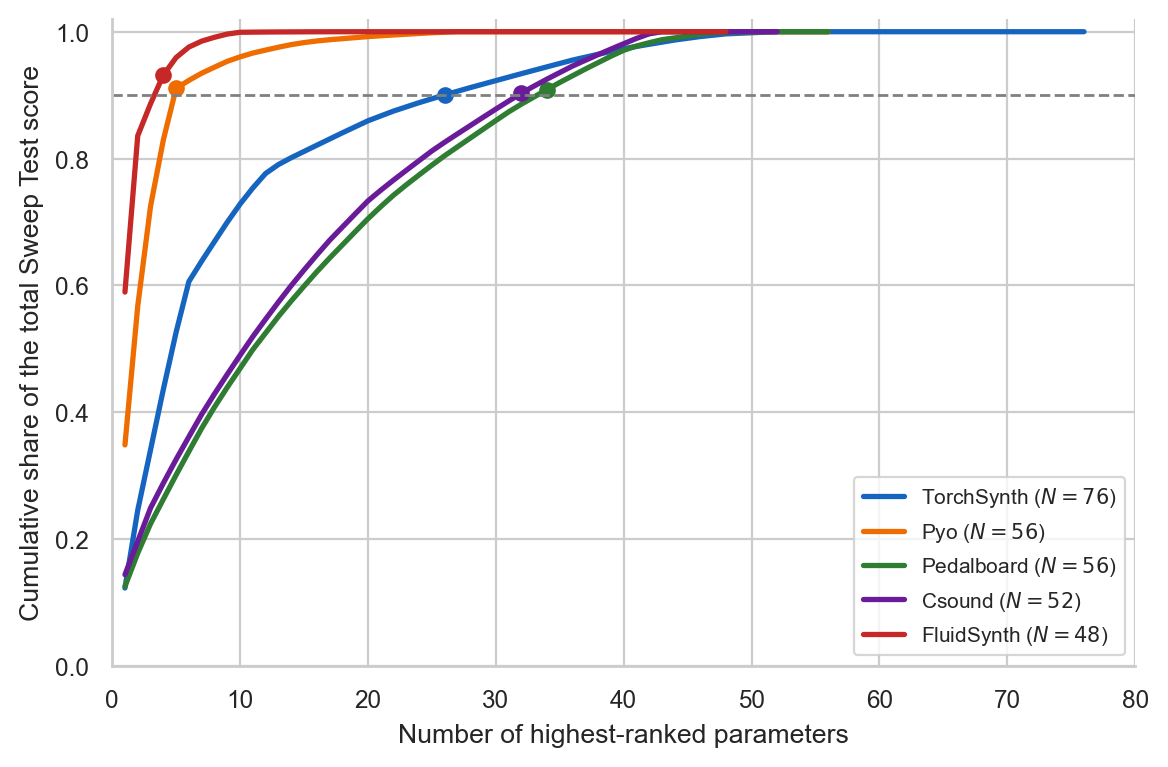}
  \caption{Cumulative share of the total Sweep Test score in rank order. Dashed line: $90\%$ of the total; markers: where each engine reaches it.}
  \label{fig:concentration}
\end{figure}

Figure~\ref{fig:concentration} shows how the measured sensitivity accumulates over the ranked parameters of each engine, and it qualifies the common assumption that a large control space is mostly inert. The assumption holds strongly for Pyo and FluidSynth, where fewer than five controls account for $90\%$ of the total measured influence, and for TorchSynth at 26 of 76; Pedalboard and Csound, however, distribute their influence far more evenly. How much of a control space can be set aside is therefore a property of the individual instrument, which is precisely why the framework measures every engine separately rather than transferring a ranking between them.

\begin{table}[t]
\centering
\caption{Top-5 parameters per backend by $S_i$. $^\dagger$Render-noise artefact (Section~\ref{sec:val_results}).}
\label{tab:top5}
\scriptsize
\setlength{\tabcolsep}{2.5pt}
\resizebox{\columnwidth}{!}{%
\begin{tabular}{@{}lllllc@{}}
\toprule
Rank & TorchSynth & Pyo & Pedalboard & Csound & FluidSynth \\
\midrule
1 & env-1 release & amp release & filter mode & filter type & mod-env sustain$^\dagger$ \\
2 & env-2 release & filter type & filter cutoff & master tune & expression \\
3 & LFO-1-env release & filter resonance & delay time & filter cutoff & stereo balance \\
4 & LFO-2-env release & delay mix & delay mix & osc-1 detune & pan \\
5 & osc-2 tuning & delay time & master tune & delay mix & pitch bend \\
\midrule
$S_{\mathrm{rank\,1}}$ & 0.92 & 0.95 & 1.00 & 1.00 & 1.00$^\dagger$ \\
\bottomrule
\end{tabular}%
}
\end{table}

\subsection{Sobol Test Rankings}
The right half of Table~\ref{tab:top15} shows the TorchSynth top-15 by $V_i$. The variance view largely reproduces the sweep view with one instructive reordering: the tuning and modulation depth of oscillator~2 rise to ranks 1--2 because pitch excursions relocate energy across many time--frequency cells at once, generating the largest per-cell variance, while the envelope releases follow immediately at ranks 3--6. On Pedalboard and Csound the variance leader matches the sweep leader outright; FluidSynth's variance list is led by the expression controller. For Pyo the two leaders swap: the filter-type selector produces the largest output variance while the amplifier release follows at rank 2—the same two controls as the sweep, in reversed order.

\begin{table}[t]
\centering
\caption{TorchSynth top-15 by the Sweep Test score $S_i$ (left) and by the Sobol Test variance $V_i$ (right).}
\label{tab:top15}\scriptsize
\setlength{\tabcolsep}{3.5pt}
\begin{tabular}{@{}rlc@{\hskip 10pt}rlr@{}}
\toprule
\multicolumn{3}{c}{by Sweep Test $S_i$} & \multicolumn{3}{c}{by Sobol Test $V_i$} \\
\cmidrule(r){1-3} \cmidrule(l){4-6}
Rank & Parameter & $S_i$ & Rank & Parameter & $V_i$ \\
\midrule
1.5  & env-1 release        & 0.92 & 1.0  & osc-2 tuning            & 491\,809 \\
1.5  & env-2 release        & 0.92 & 2.5  & env-1 release           & 436\,892 \\
3.5  & LFO-1 level env rel. & 0.72 & 2.5  & env-2 release           & 436\,892 \\
3.5  & LFO-2 level env rel. & 0.72 & 4.0  & osc-2 mod depth         & 436\,844 \\
5.0  & osc-2 tuning         & 0.68 & 5.5  & LFO-1 level env rel.    & 307\,692 \\
6.0  & osc-2 mod depth      & 0.60 & 5.5  & LFO-2 level env rel.    & 307\,692 \\
7.0  & osc-1 tuning         & 0.24 & 7.0  & mixer noise level       & 147\,049 \\
8.0  & osc-1 mod depth      & 0.23 & 8.0  & mixer osc-2 level       & 48\,961 \\
9.0  & osc-2 initial phase  & 0.23 & 9.0  & osc-2 shape             & 21\,661 \\
10.0 & osc-1 initial phase  & 0.21 & 10.0 & osc-2 initial phase     & 18\,907 \\
11.0 & mixer osc-2 level    & 0.19 & 11.0 & osc-1 mod depth         & 16\,802 \\
12.0 & mixer noise level    & 0.17 & 12.0 & osc-1 tuning            & 13\,915 \\
13.0 & mixer osc-1 level    & 0.11 & 13.0 & osc-1 initial phase     & 13\,290 \\
14.0 & osc-2 shape          & 0.08 & 14.5 & env-1 attack            & 11\,040 \\
15.5 & mod: env-1 to osc-1 amp & 0.07 & 14.5 & env-2 attack         & 11\,040 \\
\bottomrule
\end{tabular}
\end{table}

\subsection{Cross-Synthesizer Comparison}
\label{sec:crosssynth}
Comparing the five engines exposes convergent structure at the sensitive end—the three engines with a discrete filter-mode selector place it at or near the top, and the two whose tail is shaped by an amplitude envelope place the release at rank 1, since on a 3\,s clip with a 2\,s tail that stage governs more of the signal than any other control—and divergent structure at the insensitive end, where each engine's inert set follows its own architecture: modulation-matrix routings on TorchSynth, reverb geometry on Pyo, modulator-envelope and key-follow controls on Pedalboard and Csound, and the SF2 generator overrides that the Grand Piano preset renders inaudible on FluidSynth.

Parameters performing the same musical function are more revealing still. Table~\ref{tab:functional} summarises seven such families, a dash marking an engine without that module, and Fig.~S1 of the supplementary material plots them. The same function can occupy any position depending on the implementation: the amplifier-envelope release is rank 1 on TorchSynth ($S_i = 0.92$) and Pyo ($0.95$), mid-tier on Pedalboard ($0.29$), yet measures nothing on Csound, whose tail is dominated by its delay and reverb at the $0.5$ anchor, or on FluidSynth, whose piano sample decays before the release override becomes audible. Filter cutoff is rank 2--3 on Pedalboard and Csound but nearly inert on Pyo. The master-gain family, conversely, behaves identically wherever it can: after loudness conditioning pure volume controls fall to the bottom of every deterministic engine, FluidSynth's expression controller being the documented exception that reaches rank 2 solely through its mute transition. An audit of the exported corpus confirms this at the audio level—no deterministic engine ever renders a silent clip, that controller mutes 30 of its 128 excitations, and the parameters whose clips are bit-identical coincide exactly with those scored at zero variance—and the supplementary material reports it in full with the timbre-only ranking it makes possible. The lesson is that \emph{sensitivity is a property of the implementation and the operating point, not of the parameter's label}: a ranking measured on one synthesizer cannot be transferred to another, which is why the framework treats every engine as its own black box.

\begin{table}[t]
\centering
\caption{Functionally equivalent parameters across engines: $S_i$ with rank in parentheses.}
\label{tab:functional}
\scriptsize
\setlength{\tabcolsep}{3pt}
\resizebox{\columnwidth}{!}{%
\begin{tabular}{@{}lccccc@{}}
\toprule
Function & TorchSynth & Pyo & Pedalboard & Csound & FluidSynth \\
\midrule
Amp-env release & 0.92 (1) & 0.95 (1) & 0.29 (7) & 0.00 (46) & 0.00 (22) \\
Amp-env attack  & 0.06 (21) & 0.03 (9) & 0.02 (45) & 0.22 (8) & 0.00 (47) \\
Filter cutoff   & --- & 0.03 (8) & 0.42 (2) & 0.37 (3) & 0.03 (6) \\
Filter mode/type & --- & 0.60 (2) & 1.00 (1) & 1.00 (1) & --- \\
Delay mix       & --- & 0.29 (4) & 0.31 (4) & 0.26 (5) & --- \\
Master tune     & 0.68 (5) & 0.00 (22) & 0.31 (5) & 0.37 (2) & 0.00 (43) \\
Master volume   & --- & 0.03 (7) & 0.02 (44) & 0.00 (44) & 0.42 (2) \\
\bottomrule
\end{tabular}%
}
\end{table}

\subsection{Validation}
\label{sec:val_results}
Table~\ref{tab:validation} reports the Spearman correlation between the $S$ and $V$ lists. On the four engines with deterministic rendering the two independent experiments agree at $\rho = 0.961$, $0.923$, $0.874$ and $0.823$ with vanishing $p$-values: both rank the discrete filter selectors of Pedalboard and Csound first, and on TorchSynth and Pyo each test's leader sits inside the other's top two, so the ordering is a property of the synthesizer rather than of the measurement method. Substituting the symmetric aggregate $V^{\mathrm{agg}}$ for the log-STFT score leaves every verdict unchanged and strengthens two correlations (Csound $0.959$, Pedalboard $0.951$), so no conclusion rests on a single fingerprint. FluidSynth is the one weak case at $\rho = 0.319$, and its cause is identified rather than unexplained: its Sweep Test leader accumulated the engine's own render noise, and all 128 of that control's exported Sobol clips are bit-identical. Fig.~S2 of the supplementary material shows both engines parameter by parameter, labels the largest disagreements—the shortlist the framework produces for listening inspection—and dissects the FluidSynth case in full, together with Table~\ref{tab:tenparams}, which collects the ten parameters that best summarise the agreement and its exceptions.

\begin{table}[t]
\centering
\caption{Spearman $\rho$ between the $S$ and $V$ rankings at $K = 128$, for the headline log-STFT score and the five-map aggregate.}
\label{tab:validation}
\footnotesize
\setlength{\tabcolsep}{4pt}
\begin{tabular}{@{}lccccc@{}}
\toprule
Backend & $N$ & $\rho(S,V)$ & $p$-value & $\rho(S,V^{\mathrm{agg}})$ & Agreement \\
\midrule
TorchSynth  & 76 & $+0.961$ & $6.6\times10^{-43}$ & $+0.948$ & strong \\
Pedalboard  & 56 & $+0.923$ & $4.7\times10^{-24}$ & $+0.951$ & strong \\
Csound      & 52 & $+0.874$ & $2.9\times10^{-17}$ & $+0.959$ & strong \\
Pyo         & 56 & $+0.823$ & $7.2\times10^{-15}$ & $+0.820$ & strong \\
FluidSynth  & 48 & $+0.319$ & $2.7\times10^{-2}$  & $+0.303$ & weak \\
\bottomrule
\end{tabular}
\end{table}

The single weak case has an identified, mechanistic cause rather than unexplained scatter, and the FluidSynth panel of Fig.~S2 in the supplementary material shows all of it at once. FluidSynth's weak correlation is dominated by one emblematic outlier: its Sweep Test leader, the SF2 modulation-envelope sustain, shows the largest sweep deviation of the entire backend in one rendering process, yet its Sobol Test variance in another process is numerically zero. The audio corpus settles what that means—all 128 exported Sobol clips for this control are bit-identical, so its top Sweep Test score corresponds to no audio change whatever, and the deviation it accumulated in the sweep is the engine's own render noise passing a floor estimated from a single baseline pair. Below that outlier the genuinely responsive controls agree well between the tests: expression, stereo balance, pan, and pitch bend occupy the top ranks of both lists, which is why the top of that panel is as flat as the TorchSynth one beside it.

\begin{table}[t]
\centering
\caption{Ten parameters of special interest, with $S_i$ and both rank positions.}
\label{tab:tenparams}
\scriptsize
\setlength{\tabcolsep}{3pt}
\resizebox{\columnwidth}{!}{%
\begin{tabular}{@{}llllcc@{}}
\toprule
Engine & Parameter & Identifier & $S_i$ & $S$-rank & $V$-rank \\
\midrule
Pedalboard & filter mode        & \texttt{flt.mode}        & 1.00 & 1 & 1 \\
Csound     & filter type        & \texttt{flt.type}        & 1.00 & 1 & 1 \\
TorchSynth & envelope-1 release & \texttt{adsr1.release}   & 0.92 & 1 & 3 \\
TorchSynth & osc-2 tuning       & \texttt{vco2.tuning}     & 0.68 & 5 & 1 \\
Pyo        & amp-env release    & \texttt{amp.release}     & 0.95 & 1 & 2 \\
Pyo        & delay time         & \texttt{del.time}        & 0.22 & 5 & 42 \\
FluidSynth & mod-env sustain    & \texttt{gen.mod\_sustain} & 1.00 & 1 & 34 \\
FluidSynth & expression         & \texttt{cc.expression}   & 0.42 & 2 & 1 \\
TorchSynth & LFO-1-env attack   & \texttt{lfo1\_amp.attack} & 0.03 & 40 & 22 \\
Csound     & chorus rate        & \texttt{cho.rate}        & 0.14 & 20 & 4 \\
\bottomrule
\end{tabular}%
}
\end{table}

\subsection{Robustness Ablations}
\label{sec:ablations}
Four checks probe whether the rankings depend on any individual design choice; the supplementary material reports each in full. Removing each fingerprint in turn changes nothing essential: Table~\ref{tab:ablation} gives the worst case over the five removals, with the Sweep ranking holding at $\rho \ge 0.967$, the aggregate variance ranking at $\rho \ge 0.938$ and the validation itself at $\rho \ge 0.919$ on the deterministic trio. Rebuilding $S_i$ under percentile ($5$th--$95$th) and rank scaling reproduces the min--max ranking at $\rho \ge 0.975$ and $\tau \ge 0.89$ on the bit-deterministic engines, so the ordering is not an artefact of that normalisation. Removing the loudness-conditioning stage leaves the overall orderings intact ($\rho = 0.90$--$0.99$) but lifts the gain family to the top—Pedalboard's master volume moves from rank 44 to 11 in the Sweep Test and from 47 to 7 in the Sobol Test—which demonstrates the confound of Section~\ref{sec:conditioning} empirically. Finally, both tests are converged well below their operating budgets: sweep grids of $G = 26$ and $51$ reproduce the $G = 101$ ranking at $\rho = 0.998$ and $0.9995$, and a Sobol budget of $K = 32$ moves the validation correlation by at most $0.006$ on the deterministic engines, so the test could run at a quarter of its cost. FluidSynth alone moves erratically with budget, which is itself a cheap diagnostic of engine nondeterminism.

\begin{table}[t]
\centering
\caption{Robustness ablations.}
\label{tab:ablation}
\footnotesize
\setlength{\tabcolsep}{4pt}
\begin{tabular}{@{}lccccc@{}}
\toprule
& \multicolumn{3}{c}{Leave-one-out (worst of 5)} & \multicolumn{2}{c}{Scaling of $S_i$} \\
\cmidrule(lr){2-4} \cmidrule(l){5-6}
Backend & $S$ & $V^{\mathrm{agg}}$ & valid. & pctl. & rank \\
\midrule
TorchSynth  & $+0.985$ & $+0.964$ & $+0.919$ & $+0.994$ & $+0.987$ \\
Pedalboard  & $+0.991$ & $+0.988$ & $+0.932$ & $+0.991$ & $+0.983$ \\
Csound      & $+0.985$ & $+0.974$ & $+0.944$ & $+0.975$ & $+0.983$ \\
Pyo         & $+0.869$ & $+0.938$ & $+0.711$ & $+0.812$ & $+0.821$ \\
FluidSynth  & $+0.967$ & $+0.999$ & $+0.287$ & $+0.870$ & $+0.889$ \\
\bottomrule
\end{tabular}
\end{table}

\subsection{Spectrogram Evidence}
\begin{figure*}[ht]
  \centering
  \includegraphics[width=\textwidth]{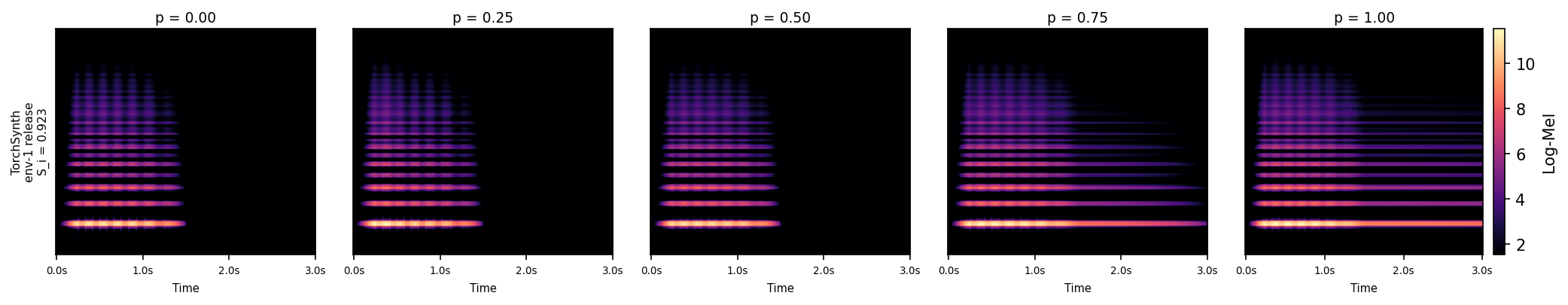}
  \caption{TorchSynth — rank-1 parameter: envelope-1 release (\texttt{adsr1.release}, $S_i{=}0.92$).}
  \label{fig:spec_torchsynth}
\end{figure*}

Fig.~\ref{fig:spec_torchsynth} shows log-mel spectrograms of TorchSynth's rank-1 parameter swept over $p \in \{0.00, 0.25, 0.50, 0.75, 1.00\}$ on conditioned audio: the envelope-1 release extends the decay tail progressively toward the full 3\,s as $p$ grows, the visual signature of the score it receives. The strips for the other four engines, discussed in the supplementary material (Figs.~S4--S7), match their quantitative story equally—including FluidSynth's, where the panel-to-panel differences are the engine's render noise rather than a parametric response.

% ============================================================
\section{Discussion}
\label{sec:discussion}
% ============================================================

Three takeaways follow from the results. Musically, the largest single-control changes a subtractive synthesizer offers are categorical rather than continuous—a filter-mode switch saturates the score on every engine that has one—while among continuous controls the envelope release dominates wherever it shapes the tail and tuning dominates the variance view, because pitch motion relocates energy across the whole time--frequency plane. The two tests formalise complementary intuitions: how far one knob takes the sound from the current patch, against how large a family of sounds it can produce.

For sound design, the ranking is an attention map. On TorchSynth 24 of 76 parameters produce no measurable change at the mid-range anchor and the 26 highest-ranked carry $90\%$ of the total measured sensitivity (Fig.~\ref{fig:concentration}), though how far a surface can be shrunk is engine-dependent—the same measurement puts the $90\%$ mark at five parameters on Pyo but 34 of 56 on Pedalboard. Whether the discarded controls are also \emph{perceptually} negligible is a listening question this study does not settle. The same map guides expansion: sensitive parameters are the productive axes for macro controls and morphing, inert ones are safe hosts for humanising randomisation.

Methodologically, the framework keeps the fingerprint weights fixed and equal, which costs adaptivity but buys rankings that are comparable across engines and across studies, with no hyperparameter to tune; the supplementary material weighs that choice against dynamic alternatives. And because the device is treated only as a black box from $[0,1]^N$ to audio, the framework transfers to any instrument or effect exposing such a control space—amplifier modeling, worked through in the supplementary material, being the case where the gain confound is strongest and the method therefore most useful.

\subsection{Limitations}
Both tests anchor the background parameters at $0.5$, so the reported scores are one-at-a-time sensitivities around a single operating region; interactions and behaviour near other anchors are not captured, a structural property of such designs \cite{SALTELLI20101508, morris1991factorial}. An anchor ablation measures the cost: repeating the Sweep Test on TorchSynth with every background parameter at $0.25$, and again at $0.75$, reproduces the mid-range ranking at only $\rho = 0.50$ and $0.58$. The parameters responsible are the bipolar modulation routings, for which $0.5$ is the neutral point, so they register nothing there and rise into the top ten once the anchor moves. These rankings therefore characterise one operating region rather than the instrument as a whole, and the study likewise uses a single pitch, velocity and note length. The silence-at-extreme effect of Section~\ref{sec:conditioning} lets gain-family parameters top the ranking of a sampled instrument through the mute transition alone, which is why the released results carry a corpus-level detector for those parameters and a parallel timbre-only ranking. The earlier asymmetry between the indices is resolved by the symmetric aggregate $V^{\mathrm{agg}}$, which leaves every verdict unchanged, and the min--max scaling is validated against percentile and rank alternatives in Section~\ref{sec:ablations}. Finally, the render-noise floor comes from a single baseline pair, which is fragile on engines with process-level nondeterminism. Pooling it over three pairs confirms both the diagnosis and the bound of the remedy: the artefact disappears—the modulation-envelope sustain falls from rank 1 to the inert block and the expression controller becomes the measured leader—yet validation lifts only to $\rho = 0.37$, because fresh process noise re-draws the low-variance tail faster than any floor estimated within one run can absorb.

% ============================================================
\section{Conclusion}
\label{sec:conclusion}
% ============================================================
This work presented a two-stage, self-validating framework for ranking synthesizer parameters by their influence on the audio output: loudness conditioning removes the gain confound, five spectral fingerprints make time-series outputs comparable, and a sweep and an independent Sobol excitation score every parameter twice, their rank correlation measuring internal consistency. Across five architecturally unrelated engines the tests agree strongly wherever rendering is deterministic and weakly, for an identified mechanistic reason, where render noise intervenes. Because the method treats each synthesizer as a black box from $[0,1]^N$ to audio, it applies unchanged to new engines at a one-time offline cost. Future work spans other pitches and anchors, multi-parameter designs that expose interactions \cite{SOBOL2001271, saltelli:2008}, perceptual and task-based validation, and alternative fingerprints \cite{gupta2022parametersensitivitydeepfeaturebased, Han_2024, he2026timberagentgramguidedretrievalexecutable}. All scripts, results, logs and figures are available at \href{https://github.com/ogulcankibar/APSR}{this GitHub repository}.

\section*{Acknowledgement}
This paper has been submitted to the Journal of the Audio Engineering Society (JAES) for peer review.

\bibliographystyle{IEEEtran}
\bibliography{refs.bib}

@misc{turian2021billionaudiosoundsgpuenabled,
      title={One Billion Audio Sounds from GPU-enabled Modular Synthesis}, 
      author={Joseph Turian and Jordie Shier and George Tzanetakis and Kirk McNally and Max Henry},
      year={2021},
      eprint={2104.12922},
      archivePrefix={arXiv},
      primaryClass={cs.SD},
      url={https://arxiv.org/abs/2104.12922}, 
}

@misc{salimi2025evaluatingsoundsimilaritymetrics,
      title={Evaluating Sound Similarity Metrics for Differentiable, Iterative Sound-Matching}, 
      author={Amir Salimi and Abram Hindle and Osmar R. Zaiane},
      year={2025},
      eprint={2506.22628},
      archivePrefix={arXiv},
      primaryClass={cs.SD},
      url={https://arxiv.org/abs/2506.22628}, 
}

@misc{he2026timberagentgramguidedretrievalexecutable,
      title={TimberAgent: Gram-Guided Retrieval for Executable Music Effect Control}, 
      author={Shihao He and Yihan Xia and Fang Liu and Taotao Wang and Shengli Zhang},
      year={2026},
      eprint={2603.09332},
      archivePrefix={arXiv},
      primaryClass={cs.SD},
      url={https://arxiv.org/abs/2603.09332}, 
}

@misc{hagiwara2022modelinganimalvocalizationssynthesizers,
      title={Modeling Animal Vocalizations through Synthesizers}, 
      author={Masato Hagiwara and Maddie Cusimano and Jen-Yu Liu},
      year={2022},
      eprint={2210.10857},
      archivePrefix={arXiv},
      primaryClass={cs.SD},
      url={https://arxiv.org/abs/2210.10857}, 
}

@misc{novello2022goalorientedsensitivityanalysishyperparameters,
      title={Goal-Oriented Sensitivity Analysis of Hyperparameters in Deep Learning}, 
      author={Paul Novello and Gaël Poëtte and David Lugato and Pietro Marco Congedo},
      year={2022},
      eprint={2207.06216},
      archivePrefix={arXiv},
      primaryClass={stat.ML},
      url={https://arxiv.org/abs/2207.06216}, 
}

@inproceedings{steinmetz2020auraloss,
    title={auraloss: {A}udio focused loss functions in {PyTorch}},
    author={Steinmetz, Christian J. and Reiss, Joshua D.},
    booktitle={Digital Music Research Network One-day Workshop (DMRN+15)},
    year={2020}
}

@article{mckay1979comparison,
 ISSN = {00401706},
 URL = {http://www.jstor.org/stable/1268522},
 author = {M. D. McKay and R. J. Beckman and W. J. Conover},
 journal = {Technometrics},
 number = {2},
 pages = {239--245},
 publisher = {[Taylor & Francis, Ltd., American Statistical Association, American Society for Quality]},
 title = {A Comparison of Three Methods for Selecting Values of Input Variables in the Analysis of Output from a Computer Code},
 urldate = {2026-05-09},
 volume = {21},
 year = {1979}
}

@inproceedings{Belanger2016,
  author    = {Bélanger, Olivier},
  title     = {Pyo, the Python DSP toolbox},
  booktitle = {Proceedings of the 24th ACM international conference on Multimedia},
  year      = {2016},
  pages     = {1214--1217},
  publisher = {ACM},
  doi       = {10.1145/2964284.2973804},
  url       = {http://ajaxsoundstudio.com/software/pyo/}
}

@misc{sobot_peter_2023_7817838,
  author       = {Sobot, Peter},
  title        = {Pedalboard},
  month        = jul,
  year         = 2021,
  publisher    = {Zenodo},
  doi          = {10.5281/zenodo.7817838},
  url          = {https://doi.org/10.5281/zenodo.7817838}
}

@inproceedings{Lazzarini2016CsoundAS,
  title={Csound: A Sound and Music Computing System},
  author={Victor Lazzarini and Steven Yi and John ffitch and Joachim Heintz and {\O}yvind Brandtsegg and I. J. McCURDY},
  booktitle={Proceedings of the Linux Audio Conference},
  year={2016},
  url={https://api.semanticscholar.org/CorpusID:114456709}
}

@techreport{itur_bs1770_5_2023,
  author      = {{International Telecommunication Union}},
  title       = {Algorithms to measure audio programme loudness and true-peak audio level},
  institution = {ITU Radiocommunication Sector},
  type        = {Recommendation},
  number      = {ITU-R BS.1770-5},
  year        = {2023},
  month       = {11},
  address     = {Geneva, Switzerland},
  url         = {https://www.itu.int/publ/R-REC/en}
}

@article{SOBOL196786,
  title   = {On the distribution of points in a cube and the approximate evaluation of integrals},
  journal = {USSR Computational Mathematics and Mathematical Physics},
  volume  = {7},
  number  = {4},
  pages   = {86--112},
  year    = {1967},
  issn    = {0041-5553},
  doi     = {10.1016/0041-5553(67)90144-9},
  url     = {https://www.sciencedirect.com/science/article/pii/0041555367901449},
  author  = {Sobol', I. M.}
}

@article{SOBOL2001271,
title = {Global sensitivity indices for nonlinear mathematical models and their Monte Carlo estimates},
journal = {Mathematics and Computers in Simulation},
volume = {55},
number = {1},
pages = {271-280},
year = {2001},
note = {The Second IMACS Seminar on Monte Carlo Methods},
issn = {0378-4754},
doi = {https://doi.org/10.1016/S0378-4754(00)00270-6},
url = {https://www.sciencedirect.com/science/article/pii/S0378475400002706},
author = {I.M Sobol'}
}

@book{saltelli:2008,
  author = {Saltelli, A.},
  description = {Global sensitivity analysis: the primer - Andrea Saltelli - Google Bücher},
  isbn = {9780470059975},
  lccn = {2007045551},
  publisher = {John Wiley},
  title = {Global sensitivity analysis: the primer},
  url = {http://books.google.at/books?id=wAssmt2vumgC},
  year = 2008
}

@article{davis1980comparison,
  title={Comparison of parametric representations for monosyllabic word recognition in continuously spoken sentences},
  author={Davis, S.B. and Mermelstein, P.},
  journal={IEEE Transactions on Acoustics, Speech, and Signal Processing},
  volume={28},
  number={4},
  pages={357--366},
  year={1980},
  publisher={IEEE}
}

@techreport{peeters2004large,
  author      = {Peeters, Geoffroy},
  title       = {A large set of audio features for sound description (similarity and classification) in the {CUIDADO} project},
  institution = {IRCAM},
  year        = {2004},
  type        = {Technical Report},
  address     = {Paris, France},
  month       = {January}
}

@article{morris1991factorial,
  author    = {Morris, Max D.},
  title     = {Factorial Sampling Plans for Preliminary Computational Experiments},
  journal   = {Technometrics},
  volume    = {33},
  number    = {2},
  pages     = {161--174},
  year      = {1991},
  doi       = {10.2307/1269043},
  url       = {https://doi.org},
  publisher = {Taylor \& Francis, Ltd.}
}

@article{SALTELLI20101508,
  title = {How to avoid a perfunctory sensitivity analysis},
  journal = {Environmental Modelling \& Software},
  volume = {25},
  number = {12},
  pages = {1508-1517},
  year = {2010},
  doi = {10.1016/j.envsoft.2010.04.012},
  author = {Saltelli, Andrea and Annoni, Paola}
}

@article{joe2008constructing,
  author    = {Joe, Stephen and Kuo, Frances Y.},
  title     = {Constructing {Sobol} Sequences with Better Two-Dimensional Projections},
  journal   = {SIAM Journal on Scientific Computing},
  volume    = {30},
  number    = {5},
  pages     = {2635--2654},
  year      = {2008},
  doi       = {10.1137/070709359},
  url       = {https://doi.org/10.1137/070709359}
}

@article{Caflisch_1998,
  title={Monte Carlo and quasi-Monte Carlo methods},
  author={Caflisch, Russel E.},
  journal={Acta Numerica},
  volume={7},
  pages={1--49},
  year={1998},
  doi={10.1017/S0962492900002804}
}

@article{spearman2010proof,
  author = {Spearman, C.},
  title = {The proof and measurement of association between two things},
  journal = {International Journal of Epidemiology},
  volume = {39},
  number = {5},
  pages = {1137--1150},
  year = {2010},
  month = {10},
  issn = {0300-5771},
  doi = {10.1093/ije/dyq191},
  url = {https://doi.org/10.1093/ije/dyq191}
}

@techreport{EBUR128_2020,
  title       = {Loudness normalisation and permitted maximum level of audio signals},
  author      = {{European Broadcasting Union (EBU)}},
  institution = {European Broadcasting Union},
  type        = {EBU Recommendation},
  number      = {R 128},
  year        = {2020},
  address     = {Geneva, Switzerland},
  url         = {https://tech.ebu.ch/docs/r/r128v4_0.pdf}
}

@misc{gupta2022parametersensitivitydeepfeaturebased,
      title={Parameter Sensitivity of Deep-Feature based Evaluation Metrics for Audio Textures}, 
      author={Chitralekha Gupta and Yize Wei and Zequn Gong and Purnima Kamath and Zhuoyao Li and Lonce Wyse},
      year={2022},
      eprint={2208.10743},
      archivePrefix={arXiv},
      primaryClass={eess.AS},
      url={https://arxiv.org/abs/2208.10743}, 
}

@article{Han_2024,
   title={Learning to Solve Inverse Problems for Perceptual Sound Matching},
   volume={32},
   ISSN={2329-9304},
   url={http://dx.doi.org/10.1109/TASLP.2024.3393738},
   DOI={10.1109/taslp.2024.3393738},
   journal={IEEE/ACM Transactions on Audio, Speech, and Language Processing},
   publisher={Institute of Electrical and Electronics Engineers (IEEE)},
   author={Han, Han and Lostanlen, Vincent and Lagrange, Mathieu},
   year={2024},
   pages={2605–2615}
}

@misc{turian2020imsorrylossspectrallybased,
      title={I'm Sorry for Your Loss: Spectrally-Based Audio Distances Are Bad at Pitch}, 
      author={Joseph Turian and Max Henry},
      year={2020},
      eprint={2012.04572},
      archivePrefix={arXiv},
      primaryClass={cs.SD},
      url={https://arxiv.org/abs/2012.04572}, 
}

@misc{vahidi2023mesostructuresspectrogramlossdifferentiable,
      title={Mesostructures: Beyond Spectrogram Loss in Differentiable Time-Frequency Analysis}, 
      author={Cyrus Vahidi and Han Han and Changhong Wang and Mathieu Lagrange and György Fazekas and Vincent Lostanlen},
      year={2023},
      eprint={2301.10183},
      archivePrefix={arXiv},
      primaryClass={cs.SD},
      url={https://arxiv.org/abs/2301.10183}, 
}

@misc{szwarcberg2025geometricsensitivitymodalparameters,
      title={Geometric sensitivity of modal parameters in wind instrument models: a case study on saxophone intonation}, 
      author={Nathan Szwarcberg and Tom Colinot and Christophe Vergez and Michaël Jousserand},
      year={2025},
      eprint={2506.16220},
      archivePrefix={arXiv},
      primaryClass={physics.class-ph},
      url={https://arxiv.org/abs/2506.16220}, 
}

@inproceedings{Sun2021,
  author    = {Xifu Sun and Barry Croke and Stephen Roberts and Anthony Jakeman},
  title     = {Investigation of determinism-related issues in the {S}obol's low-discrepancy sequence for producing sound global sensitivity analysis indices},
  booktitle = {Proceedings of the 19th Biennial Computational Techniques and Applications Conference, CTAC-2020},
  editor    = {William McLean and Shev Macnamara and Judith Bunder},
  series    = {ANZIAM J.},
  volume    = {62},
  pages     = {C84--C97},
  year      = {2021},
  month     = dec,
  doi       = {10.21914/anziamj.v62.16094},
  url       = {http://journal.austms.org.au/ojs/index.php/ANZIAMJ/article/view/16094},
  note      = {Accessed: 2021-12-09}
}

@misc{bogdan2026instrumentalautomaticsynthesizerparameter,
      title={INSTRUMENTAL: Automatic Synthesizer Parameter Recovery from Audio via Evolutionary Optimization}, 
      author={Philipp Bogdan},
      year={2026},
      eprint={2603.15905},
      archivePrefix={arXiv},
      primaryClass={cs.SD},
      url={https://arxiv.org/abs/2603.15905}, 
}

\vspace{12pt}
\end{document}